\documentclass[12pt]{article}
\usepackage{amsmath,amsfonts}
\makeatletter \@addtoreset{equation}{section}

\usepackage{cite}
\usepackage{bm}
\usepackage{dcolumn}
\makeatother
\def\one{{\hbox{ 1\kern-.8mm l}}}
\newcommand{\Dslash}{\not{\hbox{\kern-4pt $D$}}}
\newcommand{\pdslash}{\not{\hbox{\kern-2pt $\partial$}}}
\newcommand{\be}{\begin{equation}}
\newcommand{\bea}{\begin{eqnarray}}
\newcommand{\eea}{\end{eqnarray}}
\newcommand{\ba}{\begin{array}}
\newcommand{\ea}{\end{array}}
\newcommand{\ee}{\end{equation}}

\begin{document}

\begin{titlepage}
\vspace*{1mm}%
\hfill%
\vbox{
    \halign{#\hfil        \cr
           IPM/P-2009/008 \cr
                     } 
      }  
\vspace*{15mm}%
\begin{center}

{{\Large {\bf On AdS/CFT of Galilean Conformal Field Theories }}}

\vspace*{15mm} \vspace*{1mm} {Mohsen Alishahiha$^{a}$, Ali
Davody$^{a,b}$ and Ali Vahedi$^{a,b}$}

 \vspace*{1cm}
{\it ${}^a$ School of physics, Institute for Research in Fundamental Sciences (IPM)\\
P.O. Box 19395-5531, Tehran, Iran \\ }

\vspace*{.4cm}

{\it ${}^b$ Department of Physics, Sharif University of Technology \\
P.O. Box 11365-9161, Tehran, Iran}

\vspace*{.4cm} {  alishah, davody, vahedi@ipm.ir}

\vspace*{2cm}
\end{center}

\begin{abstract}
\end{abstract}
We study a new contraction of a $d+1$ dimensional relativistic
conformal algebra where $n+1$ directions remain unchanged. For
$n=0,1$ the resultant algebras admit infinite dimensional extension
containing one and two copies of Virasoro algebra, respectively. For
$n> 1$ the obtained algebra is finite dimensional containing an
$so(2,n+1)$ subalgebra. The gravity dual is provided by taking a
Newton-Cartan like limit of the Einstein's equations of AdS space
which singles out an $AdS_{n+2}$ spacetime. The infinite dimensional
extension of $n=0,1$ cases may be understood from the fact that the
dual gravities contain $AdS_2$ and $AdS_3$ factor, respectively. We
also explore how the AdS/CFT correspondence works for this case
where we see that the main role is playing by $AdS_{n+2}$ base
geometry.

\end{titlepage}

\section{Introduction}
Non-relativistic AdS/CFT correspondence has recently been studied in
several papers including \cite{Son:2008ye}
- \cite{Correa:2008bi}. Actually
non-relativistic CFTs may be obtained from relativistic CFTs by
making use of a {\it non-relativistic limit}. In general by taking a
non-relativistic limit it means that we are sending  the speed of
light to infinity. More precisely we have $v/c\rightarrow 0$ where
$v$ is the typical speed of the model. We note, however, that there
are several ways to take this limit which may  even reduce the
dimensions of the spacetime too.

To explore the procedures of taking the non-relativistic limit we
will start from a relativistic CFT in $d+1$ dimensions parametrized
by $t,x_i$ for $i=1,\cdots,d$. To proceed let us decompose the
coordinates as $(x^+,x^-,x_i),\;i=1,\cdots,d-1$, where the light
like coordinates $(x^+,x^-)$ are defined by
\begin{eqnarray}
x^+=\frac{1}{\sqrt{2}}(t+x_d),\;\;\;\;\;\;\;\;\;\;x^-=\frac{1}{\sqrt{2}}(t-x_d).
\end{eqnarray}
Next we compactify the light like coordinate $x^-$ and identify the
momentum along the light like coordinate with the number operator of
the non-relativistic CFT. Then we look for those generators of the
relativistic conformal algebra that commute with the number operator
which altogether construct an algebra. The resultant algebra is the
Schr\"odinger algebra \cite{{Hagen:1972pd},{Niederer:1972zz}} which
is the symmetry of the Schr\"odinger equation. In other words the
Schr\"odinger group may be thought of as a subgroup of $SO(2,d+1)$
with fixed momentum along the null direction (see for example
\cite{{Burdet:1977qw},{Duval:1984cj},{Henkel:2003pu},Alvarez:2009nz,{Duval:1990hj},{D3}}).
A theory with this symmetry is a non-relativistic CFT with the
following scaling symmetry \be x^+\rightarrow \lambda^2
x^+,\;\;\;\;\;\;\;\;\;x_i\rightarrow \lambda x_i. \ee Note that
starting from $d+1$ dimensional relativistic CFT the obtained theory
is a non-relativistic CFT in $d$ dimensions. This symmetry, for
example, is relevant to study cold atoms \cite{Son:2008ye}. The
generators of the corresponding algebra are spatial translations
$P_i$, rotations $M_{ij}$, time translation $H$, Galilean boosts
$B_i$, dilation $D$, number operator $N$ and special conformal
transformation $K$. The algebra has also a central extension given
by the number operator.

Recently  gravity duals of non-relativistic CFTs have been proposed
in \cite{{Son:2008ye},{Balasubramanian:2008dm}}. It has also been
shown \cite{Alishahiha} that the asymptotic symmetry algebra of  the
corresponding geometry, in any dimension, is an infinite dimensional
algebra containing one copy Virasoro algebra  compatible with the
symmetry of non-relativistic CFT \cite{Henkel}.

On the other hand one may look for a non-relativistic conformal
algebra which scales space and time in the same way \be t\rightarrow
\lambda t,\;\;\;\;\;\;\;\;\;x_i\rightarrow \lambda x_i \ee This
algebra has recently been studied in \cite{Gopakumar}(see
also\cite{Alvarez:2007fw}) where it was shown that the corresponding
algebra may be obtained from $d+1$ dimensional relativistic
conformal algebra by making use of a contraction. Since the
contraction does not change the dimension of the algebra the
resultant algebra can be thought of as the symmetry of a
non-relativistic CFT in $d+1$ dimensions. More precisely the
contraction can be defined by the scaling $t\rightarrow t,\;x_i
\rightarrow \epsilon x_i$ in the limit of $\epsilon\rightarrow 0$.
The generators of the obtained algebra are spatial translations,
$P_i$, rotations $J_{ij}$, time translation $H$, Galilean boosts
$B_i$, dilation $D$, special conformal transformation $K$ and
spatial special conformal transformation $K_i$.

It is also shown \cite{Gopakumar} that the corresponding algebra
admits an infinite dimensional extension containing one copy of
Virasoro algebra and the  bulk gravity dual is provided by a
Newtonian gravity given in terms of a non-dynamical metric but a
dynamical torsion free affine connections. The gravity background
may also be thought of as spatial $d-1$ dimensional space fibered
over an $AdS_2$. In this sense the infinite dimensional extension
may be understood from the asymptotic isometries of this $AdS_2$.

The aim of this article is to extend the above considerations for a
new contraction in which we scale $d-n-1$ directions by $\epsilon$
and $n+1$ directions remain unchanged. Then we consider the limit of
 $\epsilon\rightarrow 0$.
The obtained algebra which we call it {\it semi-Galilean algebra}
can be thought of as a symmetry of non-relativistic CFT in $d+1$
dimensions\footnote{We note, however, that calling this theory a
non-relativistic CFT is somehow misleading as it has relativistic
properties in those directions which remained unchanged}.

We show that for $n=1$ the corresponding algebra admits an infinite
dimensional extension containing two copies of Virasoro algebra. The
corresponding gravity dual is defined on a geometry which is a $d-2$
dimensional spatial space fibered over an $AdS_3$ and, indeed, the
infinite dimensional extension can be associated to the asymptotic
isometries of $AdS_3$.

For $n\geq 2$ the contraction leads to an algebra which has
$so(2,n+1)\times so(d-n-1)$ subalgebra and the corresponding gravity
dual is defined by a geometry which is a $d-n-1$ dimensional spatial
space fibered over an $AdS_{n+2}$. The subalgebra can, then, be
identified with the isometries of $AdS_{n+2}$. Since the asymptotic
symmetry algebra of $AdS_{n+2}$ space for $n\geq 2$ is finite
dimensional, the corresponding semi-Galilean algebra  is also finite
dimensional.

The paper is organized as follows. In the next section we study the
Galilean algebra for arbitrary $n$. In section three we explore how
the AdS/CFT correspondence  works in this context. The last section
is devoted to discussions.

\section{General contraction of conformal algebra}

In this section we study non-relativistic limit of relativistic
conformal algebra in $d+1$ dimensions by making use of a
contraction. To proceed we consider the following general scaling
\be\label{scale} t\rightarrow t,\;\;\;\;\;\;y_\alpha\rightarrow
y_\alpha,\;\;\;\;\;\;x_i\rightarrow \epsilon x_i, \ee where
$\alpha=1,\cdots, n$ and $i=n+1,\cdots d$. The contraction is
defined by the above scaling in the limit of $\epsilon\rightarrow
0$. For $n=0$ this has been studied in \cite{Gopakumar} where it was
shown that the resultant contracted algebra admits an infinite
dimensional extension containing one copy of Virasoro algebra. In
what follows we would like to extend this consideration for general
$n$.

\subsection{Field theory description}

We start from a CFT in $d+1$ dimensions. The theory is invariant
under the action of generators of conformal algebra given by
rotations $J_{\mu\nu}$, translations $P_\mu$, dilation $D$ and
special conformal transformations $K_\mu$ whose representations as a
vector field acting on the $d+1$ dimensional Minkowski space are
given by \be
J_{\mu\nu}=-(x_\mu\partial_\nu-x_\nu\partial_\mu),\;\;\;\;P_\mu=\partial_\mu,\;\;\;\;D=-(x\cdot\partial),
\;\;\;\;K_\mu=-(2x_\mu (x\cdot \partial)-(x\cdot x) \partial_\mu)
\ee with $\mu=0,\cdots,d$. The aim is to contract the conformal
algebra generated by the above generators. To do this we will
consider the general scaling \eqref{scale} in the limit of
$\epsilon\rightarrow 0$. To be specific we will consider the case of
$n=1$. It is, of course, straightforward to generalize it for
arbitrary $n$.

For $n=0$ under the above rescaling and in the of
$\epsilon\rightarrow 0$ the $d+1$ dimensional conformal algebra
reduces to Galilean conformal algebra studied in \cite{Gopakumar}.
For $n=1$ the generators of the corresponding algebra as a vector
field acting on a $d+1$ dimensional Minkowski space ( for $d\geq 2$)
are given by \bea
&&J_{ij}=-(x_i\partial_j-x_j\partial_i),\;\;\;\;\;J_{i}=t\partial_i,\;\;\;\;\;
{\tilde J}_i=-y\partial_i,\;\;\;\;\;P_i=\partial_i,
\;\;\;\;\;K_i=(y^2-t^2)\partial_i,\cr &&\cr
&&D=-(x_i\partial_i+y\partial_y+t\partial_t),\;\;\;\;\;\;\;\;\;\tilde{D}=t\partial_y+y\partial_t,\;\;\;\;\;\;\;\;
P=\partial_t,\;\;\;\;\;\;\;\;\tilde{P}=\partial_y, \cr &&\cr
&&K=(t^2+y^2)\partial_t+2ty\partial_y+2tx_i\partial_i,\;\;\;\;\;\;\;
\tilde{K}=-(t^2+y^2)\partial_y-2ty\partial_t-2yx_i\partial_i. \eea
To have an insight how the semi-Galilean conformal algebra for $n=1$
could be, it is useful to define new coordinates $u=t+y,\;v=t-y$ by
which the above generators may be recast to the following
form\footnote{For example $H=\tilde{P}+P,\;C=K-\tilde{K}$ and
$E=\tilde{D}-D$.} \bea
&&H=2\partial_u,\;\;\;\;\;\;\;\;\;E=2(u\partial_u+\frac{1}{2}x_i\partial_i),\;\;\;\;\;\;\;\;\;\;
C=2(u^2\partial_u+ux_i\partial_i),\cr &&\cr &&{\bar
H}=2\partial_v,\;\;\;\;\;\;\;\;\;{\bar
E}=-2(v\partial_v+\frac{1}{2}x_i\partial_i),\;\;\;\;\;\;\;\; {\bar
C}=2(v^2\partial_v+vx_i\partial_i),\cr &&\cr
&&J_{ij}=-(x_i\partial_j-x_j\partial_i),\;\;\;P_i=\partial_i,\;\;\;B_{i}=-u\partial_i,\;\;\;
{\bar B}_i=v\partial_i,\;\;\;K_i=-uv\partial_i. \eea It is easy to
see that $(H,E,C)$ and $(\bar{H},\bar{E},\bar{C})$ generate two
copies of $SL(2,R)$ algebra. In fact to write the explicit form of
the commutation relations of the algebra it is useful to define
$L_{\pm 1, 0}, \bar{L}_{\pm 1, 0}$ and  $M_{i\; rs}\;r,s=0,1$ as
follows \bea\label{uu}
&&\{L_{-1}=\frac{1}{2}H,\;\;L_0=\frac{1}{2}E,\;\;L_1=\frac{1}{2}C\},\;\;\;\{\bar{L}_{-1}
=\frac{1}{2}{\bar
H},\;\;\bar{L}_0=-\frac{1}{2}\bar{E},\;\;\bar{L}_1=\frac{1}{2}\bar{C}\},\cr
&&\cr
&&\{M_{i\;00}=-P_i,\;\;M_{i\;01}=B_i,\;\;M_{i\;10}=-\bar{B}_i,\;\;M_{i\;11}=K_i\}.
\eea Using this notation the non-zero commutation relations of the
algebra are \bea
&&[L_n,L_m]=(n-m)L_{n+m},\;\;\;\;\;\;\;\;\;\;\;\;\;\;\;\;\;\;\;\;\;
[\bar{L}_n,\bar{L}_m]=(n-m)\bar{L}_{n+m},\cr &&\cr
&&[L_n,M_{i\;rs}]=\left(\frac{n+1}{2}-r\right)M_{i\;(n+r)s},\;
\;\;\;\;[\bar{L}_n,M_{i\;rs}]=\left(\frac{n+1}{2}-s\right)M_{i\;s
(n+s)}, \cr &&\cr
&&[M_{l\;rs},J_{ij}]=\left(\delta_{jl}M_{i\;rs}-\delta_{il}M_{j\;rs}\right),
\;\;\;\;\;\;\;\;\;[J_{ij},J_{i'j'}]=so(d-1), \eea which make the two
$SL(2,R)$ subalgebras manifest. Here $n,m=\pm 1,0$ and $r,s=0,1$.
Actually it is useful to re-express the generators of \eqref{uu} in
the following instructive closed form \bea\label{Infboun}
L_n=u^{n+1}\partial_u+\frac{n+1}{2}u^nx_i\partial_i,\;\;\;
\bar{L}_n=v^{n+1}\partial_v+\frac{n+1}{2}v^nx_i\partial_i,\;\;\;
M_{i\;rs}=-u^rv^s\partial_i. \eea From these expressions it is
natural to define the above vector fields for arbitrary integers $n$
and $r$. Indeed defining \be
J_{ij\;nm}=-u^nv^m(x_i\partial_j-x_j\partial_i), \ee one finds an
infinite dimensional algebra as follows \bea
&&[L_n,L_m]=(n-m)L_{n+m},\;\;\;\;\;\;\;\;\;\;\;\;\;\;\;\;\;\;\;\;\;\;\;
[\bar{L}_n,\bar{L}_m]=(n-m)\bar{L}_{n+m},\cr &&\cr
&&[M_{i\;nm},M_{j\;n'm'}]=0,\;\;\;\;\;\;\;\;\;\;\;\;\;\;\;\;\;\;\;\;\;\;\;\;\;\;\;\;\;\;\;[L_n,\bar{L}_m]=0,
\cr &&\cr
&&[L_n,M_{i\;ml}]=\left(\frac{n+1}{2}-m\right)M_{i\;(n+m)l},\;
\;\;\;\;[\bar{L}_n,M_{i\;ml}]=\left(\frac{n+1}{2}-l\right)M_{i\;m
(n+l)} \cr &&\cr
&&[L_n,J_{ij\;ml}]=-mJ_{ij\;(n+m)l},\;\;\;\;\;\;\;\;\;\;\;\;\;\;\;\;\;\;\;\;\;\;
[\bar{L}_n,J_{ij\;ml}]=-lJ_{ij\;m (n+l)}\cr &&\cr
&&[M_{l\;nm},J_{ij\;n'm'}]=\left(\delta_{jl}M_{i\;(n+n')(m+m')}-\delta_{il}M_{j\;(n+n')(m+m')}\right).
\eea Moreover the $J_{ij\;nm}$'s generate an $so(d-1)$ affine
algebra.

As an conclusion we observed that the (semi) Galilean conformal
algebra obtained from the relativistic conformal algebra using the
contraction \eqref{scale} admits an infinite dimensional extension
for $n=0,1$. This may be understood from the fact that in these
cases there is at least an $SL(2,R)$ subalgebra which may be
extended to a Virasoro algebra. As we will see from gravity point of
view this corresponds to the fact that in these cases the gravity
dual develops  $AdS_2$ or $AdS_3$ geometries for $n=0$ and $n=1$,
respectively.

Actually the above procedure may be generalized for $n\geq 2$ where
we will get an algebra containing an $so(2,n+1)\times so(d-n-1)$
subalgebra. We note, however, that in this case the resultant
semi-Galilean algebra does not admit an infinite dimensional
extension. As we will see in the next section the reason may be
understood from the fact that in this case the gravity background
develops an $AdS_{n+2}$ geometry which has finite dimensional
asymptotic symmetry algebra.

\subsection{Gravity description}

Following AdS/CFT correspondence we would expect that a $d+1$
dimensional CFT may have a gravity dual defined on a background
containing an $AdS_{d+2}$ factor where the CFT lives on the boundary
of the $AdS$ space. Therefore one should be able to take the
non-relativistic limit from both side of the duality. In particular
we would like to carry out the contraction of the previous section
on the AdS part of the bulk geometry.

To proceed consider the metric of an $AdS_{d+2}$ space in the
Poincar\'e coordinates \be
ds^2=\frac{-dt^2+dy^2_\alpha+dx_i^2+dz^2}{z^2}. \ee The
non-relativistic limit of the previous section can be generalized to
the bulk geometry as follows \be\label{AdSscaling} t\rightarrow
t,\;\;\;\;\;z\rightarrow z,\;\;\;\;\; y_\alpha\rightarrow
y_\alpha,\;\;\;\;x_i\rightarrow \epsilon x_i. \ee In the limit of
$\epsilon\rightarrow 0$ where only $t,z$ and $y_\alpha$ survive the
contraction the resultant geometry develops an $AdS_{n+2}$ space.
The rest $d-n$ dimensional space parametrized by $x_i$ are fibered
over the $AdS_{n+2}$ base spacetime. Indeed as it was argued in
\cite{Gopakumar} the corresponding gravity dual should be given in
terms of the Newton-Cartan like description where $AdS_{n+2}$ plays
the special role of the time. In this formalism the metric is
non-dynamical and the dynamics are given by torsion free affine
connections. More precisely following \cite{Gopakumar} one may
define a contravariant tensor $\gamma=\gamma^{MN}\partial_M\otimes
\partial_N$ with $M,N=\{t,z,\alpha,i\}$. It has $n+2$ zero
eigenvalues  corresponding to $\{t,z,y_\alpha\}$ which parametrize
the base $AdS_{n+2}$ space with the metric \be\label{AdSn}
ds^2=g_{ab}dx^adx^b=\frac{-dt^2+dy^2_\alpha+dz^2}{z^2}. \ee The
affine connections $\Gamma^M_{NL}$ are compatible with both base AdS
space as well as the $d-n$ dimensional spatial fiber \be
\nabla_M\gamma^{ML}=0,\;\;\;\;\;\nabla_Mg_{ab}=0. \ee In our case
the dynamical connection may be given by
$\Gamma^i_{ab}=\partial_i\Phi_{ab}$ \cite{Gopakumar}.

Following the general lore of the AdS/CFT correspondence
\cite{Maldacena:1997re} we would expect that if the above Newtonian
gravity provides a gravity dual of the semi-Galilean conformal filed
theory one should be able to see the semi-Galilean symmetry algebra
as the asymptotic symmetry algebra of the above geometry in the
sense of Brown and Henneaux construction \cite{Brown:1986nw}. In
what follows we will show that this is, indeed, the case. To be
specific we will consider the case of $n=1$ which turns out to be
more interesting case. Generalization to other cases is
straightforward.

In the Poincar\'e coordinates the Killing vectors of $AdS_{d+2}$ are
given by \bea
&&J_{\mu\nu}=-(x_\mu\partial_\nu-x_\nu\partial_\mu),\;\;\;\;\;\;\;\;\;\;\;\;\;\;\;\;\;\;\;
D=-(x^\mu\partial_\mu+z\partial_z),\cr &&\cr &&K_\mu=-(2x_\mu (x^\nu
\partial_\nu+z\partial_z)-(x^\nu x_\nu+z^2)
\partial_\mu),\;\;\;\;\;\;\;P_\mu=\partial_\mu. \eea Using the
scaling \eqref{AdSscaling} the resultant contracted Killing vectors
read \bea
&&P_i=\partial_i,\;\;\;\;B_i=t\partial_i,\;\;\;\;\tilde{B}_i=-y\partial_i,
\;\;\;\;{K}_i=(t^2-y^2-z^2)\partial_i, \cr &&\cr
&&D=-(t\partial_t+y\partial_y+x_i\partial_i+z\partial_z),\;\;\;\;\;\tilde{D}
=t\partial_y+y\partial_t,\;\;\;\;J_{ij}=-(x_i\partial_j-x_j\partial_i),\cr
&&\cr &&K=-(t^2+y^2+z^2)\partial_t-2zt\partial_z-2ty\partial_y-2t
x_i\partial_i,\;\;\;\;\;\;P=\partial_t,\cr &&\cr
&&\tilde{K}=(t^2+y^2-z^2)\partial_y+2ty\partial_t+2zy\partial_z+2yx_i\partial_i,\;\;\;\;\;\tilde{P}=\partial_y.
\eea Note that to make the comparison more transparent we have used
the same labeling for bulk and boundary generators. Following our
previous discussions setting $u=t+y, v=t-y$ the above Killing
vectors may be recast to the following form \bea
&&H=2\partial_u,\;\;\;\;\;\;\;\;E=2(u\partial_u+\frac{1}{2}x_i\partial_i+\frac{1}{2}z\partial_z),\;\;\;\;\;\;\;\;\;
C=2(u^2\partial_u+u(x_i\partial_i+z\partial_z)+z^2\partial_v),\cr
&&\cr &&{\bar H}=2\partial_v,\;\;\;\;\;\;\;\;{\bar
E}=-2(v\partial_v+\frac{1}{2}x_i\partial_i+\frac{1}{2}z\partial_z),\;\;\;\;\;\;\;
{\bar
C}=-2(v^2\partial_v+v(x_i\partial_i+z\partial_z)+z^2\partial_u),\cr
&&\cr
&&J_{ij}=-(x_i\partial_j-x_j\partial_i),\;\;\;P_i=\partial_i,\;\;\;B_{i}=u\partial_i,\;\;\;
{\bar B}_i=v\partial_i,\;\;\;K_i=(uv-z^2)\partial_i. \eea which
reduce to those in the previous section in the limit of
$z\rightarrow 0$ where we approach the boundary of the $AdS_3$.

Let us define infinite dimensional vector fields in the bulk as
follows \bea
L_n&=&u^{n+1}\partial_u+\frac{n+1}{2}u^n(x_i\partial_i+z\partial_z)+\frac{n(n+1)}{2}u^{n-1}z^2\partial_v,\cr
&&\cr
\bar{L}_n&=&v^{n+1}\partial_v+\frac{n+1}{2}v^n(x_i\partial_i+z\partial_z)+\frac{n(n+1)}{2}v^{n-1}z^2\partial_u,\;
\eea which can be properly identified with $H,E,C$ and
$\bar{H},\bar{E},\bar{C}$ for $n=\pm 1,0$ which at the boundary
where $z\rightarrow 0$ reduce to those in \eqref{Infboun}. It is
interesting to note that these vector fields asymptotically obey two
copies of Virasoro algebra, {\it i.e.} \be
[L_n,L_m]=(n-m)L_{n+m},\;\;\;\;[\bar{L}_n,\bar{L}_m]=(n-m)\bar{L}_{n+m},\;\;\;\;[\bar{L}_n,L_m]={\cal
O} (z^4). \ee The action of the Virasoro generators on the metric of
the base manifold, $AdS_3$, is given \bea L_n\;:\;ds^2\rightarrow
ds^2+\frac{n(n^2-1)}{2}u^{n-2}du^2,\;\;\;\; \bar{L}_n\;:\;
ds^2\rightarrow ds^2+\frac{n(n^2-1)}{2}v^{n-2}dv^2. \eea Therefore
the generators of two $SL(2,R)$'s given by $L_{\pm,0}$ and
$\bar{L}_{\pm,0}$ are the exact isometries of the base metric, as
expected, while for other $n$'s they generate the asymptotic
symmetry which preserve the following boundary conditions \be
\left(\begin{array}{lll} h_{uu}={\cal O}(1)&h_{uv}={\cal
O}(1)&h_{uz}={\cal O}(z)\cr h_{vu}=h_{uv}&h_{vv}={\cal
O}(1)&h_{vz}={\cal O}(z)\cr h_{zu}=h_{uz}&h_{zv}=h_{vz}&h_{zz}={\cal
O}(1)
\end{array}
\right)\ . \ee On the other hand requiring to have asymptotically a
closed algebra we will have to extend the other generators as
follows \be
M_{i\;nm}=-(u^nv^m-nmu^{n-1}v^{m-1}z^2)\partial_i,\;\;\;\;J_{ij\;nm}=-u^nv^m(x_i\partial_j-x_j\partial_i),
\ee which for $m,n=0,1$ can be identified with $P_i, B_i,\bar{B}_i,
K_i$ and at the boundary where $z\rightarrow 0$ reduce to that in
\eqref{Infboun}. It is easy to see that \bea
&&[L_n,M_{i\;lm}]=\left(\frac{n+1}{2}-l\right)M_{i\;(n+l)m}+{\cal
O}(z^4),\;\;\;\;[L_n,J_{ij\;lm}]=-lJ_{ij\;(n+l)m}+ {\cal O}(z^2),\cr
&&\cr &&[\bar{L}_n,M_{i\;lm}]=\left(\frac{n+1}{2}-m\right)M_{i\;l
(n+m)}+{\cal O}(z^4),\;\; [\bar{L}_n,J_{ij\;lm}]=-mJ_{ij\;l(n+m)}+
{\cal O}(z^2),\cr &&\cr
&&[M_{l\;nm},J_{ij\;n'm'}]=\bigg(\delta_{jl}M_{i\;(n+n')(m+m')}-\delta_{il}M_{j\;(n+n')(m+m')}\bigg)
+{\cal O}(z^2),\cr &&\cr &&[M_{i\;nm},M_{j\;n'm'}]=0.\;\; \eea As a
conclusion we have demonstrated that the asymptotic symmetry algebra
of our bulk geometry is the semi-Galilean conformal algebra studied
in the previous section.

It is straightforward to generalize the above considerations for
$n\geq 2$ where the base space will be $AdS_{n+2}$. In this case
from the base space we find an $so(2,n+1)$ factor while from the
fiber one gets an $so(d-n-1)$ subalgebra which is compatible with
those studies in the previous section. We note that for the case of
$n\geq 2$ the semi-Galilean conformal algebra is finite dimensional
due to the fact the the asymptotic symmetry of $AdS_{n+2}$ for
$n\geq 2$ is finite dimensional.

\section{AdS/CFT description of theory with Galilean conformal symmetry}

In this section we would like to study the AdS/CFT correspondence
for the Galilean conformal field theory. Following the relativistic
CFT one would expect that in the Galilean CFT the asymptotic states
cannot be defined and the physical observables would be correlation
functions. Therefore the task is to compute $N$-point functions of
operators in the Galilean CFT which is the aim of this section.

In what follows we will mainly consider the case of $n=0$, though
the procedure may be generalized for other $n$'s.

\subsection{Field theory description}

Consider a Galilean CFT in $d+1$ dimensions. As we have seen the
corresponding algebra can be obtained from the relativistic CFT by a
contraction. Therefore one may naively expect that the $N$-point
functions of Galilean CFT can also be obtained from those in the
relativistic CFT by making use of the same contraction. For example
consider two point function of an operator $\phi$ with scaling
dimension $\Delta$ in a relativistic CFT in $d+1$ dimensions
parametrized by $t$ and $x_i$ \be \langle
\phi(t_1,x_i)\phi(t_2,y_i)\rangle
\sim\frac{1}{(-(t_1-t_2)^2+(x_i-y_i)^2)^\Delta}. \ee Using the
scaling limit \eqref{scale} and in the limit of $\epsilon
\rightarrow 0$ the two point function of the Galilean CFT reads \be
\langle \phi(t_1,x_i)\phi(t_2,y_i)\rangle \sim
\frac{1}{(t_1-t_2)^{2\Delta}}. \ee Similarly we can extend the above
procedure to $N$-point function to conclude that in general the
$N$-point function of Galilean CFT  depends only on time.

We, note, however that although the above results seem reasonable,
the way we reach the conclusion may not be correct in general. The
reason is due to the fact that the representations  of an algebra
under a contraction do not necessarily lead to  faithful (bijective)
representations \cite{Inonu:1953sp}. In other words although the
$N$-point functions we obtain by this method satisfy the Ward
identity of the Galilean CFT, it is not clear that the general form
of the $N$-point functions can be obtained from this method.
Therefore it would be interesting to evaluate the $N$-point function
of Galilean CFT directly. To do this we utilize the Ward identity of
the Galilean CFT.

The representation of the generators of the  Galilean conformal
algebra acting on an operator with dimension $\Delta$ is given
\bea\label{glo} &&J_{ij}= -(x_i\ p_j-x_j\ p_i),\;\;\; \qquad
P_0=H=-\partial _{t},\;\;\qquad P_i=\partial{_i} \qquad B_i=
t\partial{_i},
 \cr &&\cr && D =-(x_i\partial_{i} + t \partial_{t})-\Delta, \qquad  K=-(2t x_{i}
\partial_{i} +t^2 \partial_{t})-2\Delta t, \qquad K_i=t^2\partial_{i}.
\eea Note that although the Galilean conformal algebra admits an
infinite dimensional extension, we would expect that the vacuum is
only invariant under the global part of the algebra given by the
above generators. Therefore the Wrad identity of the Galilean CFT
may be written as follows
\begin{equation}\label{wa}
    \sum_i\langle 0| \phi(x_{1})\ldots Q\phi(x_{i})\ldots\phi(x_{N})|0 \rangle=0
\end{equation}
where $|0\rangle$ is a vacuum which is invariant under the global
part of the algebra. $Q\phi(x_{i})$ is the representation of   an
operator $Q$  on the field $\phi(x_{i})$ with $Q$ stands for one of
the generators in \eqref{glo}. By making use of the equation
\eqref{wa} one can write the Ward identities for $N$-point function
$G_{N}(x_{1},t_{1},\ldots,x_{N},t_{N})$. To write the explicit form
of the Wrad identities for $N$-point functions it is useful to
define new variables $t_{i2}=t_i-t_2, x_{i2}=x_i-x_2$ and
$\tilde{t}_{12}=t_1+t_2, \tilde{x}_{12}=x_1+x_2$ for $i=1,3,4\cdots,
N$.

From the Ward identities for space and time translations one finds
that $G_{N}$ depends only on $t_{i2}$ and $x_{i2}$, {\it i.e.}
$G_N(t_{12},t_{32},\cdots;x_{12},x_{32},\cdots)$. On the other hand
from $K_{i}$ one finds \bea\label{Ki}
&&[t_{12}D_{12}+t_{32}D_{32}+\ldots+t_{N2}D_{N2}]G_{N}=0,\cr
&&[t_{32}(t_{32}-t_{12})D_{32}+\ldots+t_{N2}(t_{N2}-t_{12})D_{N2}]G_{N}=0,
\eea where $D_{i2}=\frac{\partial}{\partial x_{i2}}$. From the
dilatation we get \be\label{D}
    [x_{12}D_{12}+\ldots+x_{N2}D_{N2}+t_{12}\partial_{12}+
    \ldots+t_{N2}\partial_{N2}+\lambda_{1}+\ldots+\lambda_{N}]G_{N}=0,
\ee and $K$ leads to the following differential equation
\bea\label{K}
&&[(2t_{32}x_{32}-t_{32}x_{12}-t_{12}x_{32})D_{32}+\ldots+(2t_{N2}x_{N2}-t_{N2}x_{12}-t_{12}x_{N2})D_{N2}\cr
&&\cr
&&+t_{32}(t_{32}-t_{12})\partial_{32}+\ldots+t_{N2}(t_{N2}-t_{12})\partial_{N2}
+t_{12}(\lambda_{1}-\lambda_{2}-\ldots-\lambda_{N})\cr &&\cr &&
+2\lambda_{3}t_{32}+\ldots+2\lambda_{N}t_{N}]G_{N}=0. \eea

Now the task is to solve these equations to find $N$-point
functions. We note, however, that these equations cannot fix the
$N$-point functions completely for arbitrary $N$. This is of course
the case even for relativistic one where the $N$-point functions can
be found up to unknown functions. Let us give the explicit form of
two and there point functions.
\subsubsection*{Two-point function}
From (\ref{Ki}) it is clear that the two-point function does not
depend on $x_{12}$  and from  (\ref{D})  we get
\begin{equation}\label{2p}
   G_2:= \langle\phi_1(t_1,{x_{i}}_1)\phi_2(t_2,{x_i}_2)\rangle=C\;\;
t_{12}^{-\Delta},\qquad\qquad\Delta=\Delta_{1}+\Delta_{2}
\end{equation}
where $C$ is a constant.
\subsubsection*{Three-point function}
From (\ref{Ki}) it is clear that the three-point function does not
depend on $x_{12}$ and $x_{32}$ and from (\ref{D}) and (\ref{K}) one
finds
\begin{equation}\label{3p}
G_3:=\langle\phi_1(t_1,{x_{i}}_1)\phi_2(t_2,{x_i}_2)\phi_3(t_3,{x_i}_3)\rangle
={C} (\frac{1}{t_{12}})^{\Delta_{1}+\Delta_{2}-\Delta_{3}}
(\frac{1}{t_{32}})^{-\Delta_{1}+\Delta_{2}+\Delta_{3}}(\frac{1}{t_{13}})^{\Delta_{1}+\Delta_{3}-\Delta_{2}}
\end{equation}
\subsubsection*{$N$-point function}
In principle one could proceed to compute $N$-point function for
arbitrary $N$, though here we will not do that. The only comment we
would like to make is that utilizing the Ward identities one can
show that the $N$-point function depends only on $t_{i2}$'s.

\subsection{Gravity description}

In this subsection we would like to see how the $N$-point functions
we have considered in the previous section can be obtained from
gravity description. The procedure in the relativistic AdS/CFT
correspondence is to evaluate the bulk action on a classical
solution with a given boundary condition. Since for Galilean CFT the
gravity description is given in terms of the Newtonian gravity the
above description may not be directly applied in this case. To
explore the procedure we start from a propagating field on an AdS
geometry and impose the contraction we have introduced in the
previous section.

To proceed, for simplicity,  we consider a massive scalar field on
the $AdS_{d+2}$ background whose equation of motion is given by \be
\frac{1}{\sqrt{G}}\partial_M\bigg(\sqrt{G}G^{MN}\partial_N\phi(t,z,x_i)\bigg)-m^2\phi(t,z,x_i)=0,
\ee where $G_{MN}$ is the metric of the AdS geometry. To be specific
we consider the AdS geometry in the Poincar\'e coordinates
parametrized by $t,z, x_i$. Under the scaling \eqref{AdSscaling} one
has \be G_{MM}\rightarrow G_{MN},\;\;\;\;\;\partial_t\rightarrow
\partial_t,\;\;\;\;\;\partial_z\rightarrow \partial_z,
\;\;\;\;\;\partial_i\rightarrow \epsilon^{-1}\partial_i. \ee so that
\be
\left[\frac{1}{\sqrt{G}}\partial_a\bigg(\sqrt{G}g^{ab}\partial_b\phi(t,z,x_i)\bigg)-m^2\phi(t,z,x_i)\right]
+\frac{z^2}{\epsilon^2}\partial_i^2\phi(t,z,x_i)=0. \ee Here
$g_{ab}$ is the metric of $AdS_2$ base geometry. In order to have a
well behaved equation in the limit of $\epsilon\rightarrow 0$ one
should impose \be
\frac{1}{\sqrt{G}}\partial_a\bigg(\sqrt{G}g^{ab}\partial_b\phi(t,z,x_i)\bigg)-m^2\phi(t,z,x_i)=0,\;\;\;\;\;
\partial_i^2\phi(t,z,x_i)=0.
\ee The first equation may be obtained from a two dimensional action
given by \be I=\int dt dz
\;\sqrt{G}\;\frac{1}{2}\bigg(g^{ab}\partial_a\phi\partial_b\phi+m^2\phi^2\bigg),
\ee while the second equation may be treated as a constraint.
Therefore the most general solution of the equation of motion of the
above action is \be
 \phi(t,z)= z^{\frac{d+1}{2}}\; e^{-i\omega t}(A I_\alpha(\omega  z)+B K_\alpha(\omega z)),
\ee where $\alpha =\sqrt{\frac{(d+1)^{2}}{4}+m^2}$. Since in the
present case the constraint decouples from the equation of motion,
it leads to an overall factor which could depends on
$x_i$\footnote{We note, however, that the over all factor could
parametrically be divergent due to the integration over boundary
term. This might be observed by redefinition of the boundary
operators by making use of a regularization. The similar behavior
happens in the non-relativistic CFT studied in
\cite{{Balasubramanian:2008dm},{Akhavan:2009ns}}.}. It is then
straightforward to follow the general role of the AdS/CFT
correspondence to find the bulk solution by given a boundary value
as follows
\begin{equation}
\label{bbulk}
  \phi(t,z) = c \delta^{\Delta-d-1} \int dt'  \phi_\delta(t')
   \left( \frac{z}{z^2 +|t-t'|^2} \right)^\Delta,
  \end{equation}
where $\Delta=\frac{d}{2}+\alpha$ and $\phi_\delta$ denotes the
Dirichlet boundary value at $z=\delta$. This can be used to read the
two point function as follows
\begin{equation}
\label{field:2point}
   \langle \mathcal{O}(t_1) \mathcal{O}(t_2) \rangle \sim
   \frac{1}{({t_1-t_2})^{2\Delta}}.
\end{equation}
in agreement with \eqref{2p} for $\Delta_1=\Delta_2=\Delta$.

To find $N$-point function we  should add an interaction term
$\lambda_N \phi^N$ to the action. Then following the standard
AdS/CFT procedure one arrives at (see for example \cite{AdS1})
\begin{equation}
\label{field:int}
  I_N(t_1,...,t_N) \sim \int dtdz\,
  \frac{z^{-(d+2)+N\Delta}}{\left[
  \left(z^2+({t-t_1})^2\right) \ldots
  \left(z^2+({t-t_N})^2\right) \right]^\Delta}.
\end{equation}
In particular for $N=3$ we get \be \label{field:3point}
  \langle \mathcal{O}(t_1) \mathcal{O}(t_2)
  \mathcal{O}(t_3) \rangle \sim
  -\frac{\lambda_3\Gamma\left(\frac12\Delta+\alpha\right)}{2\pi^{d+1}}
  \left[\frac{\Gamma\left(\frac12\Delta\right)}{\Gamma(\alpha)}\right
  ]^3 \frac{1}{\left(t_{12} t_{31} t_{23}\right)^\Delta}.
\ee in agreement with  (\ref{3p}) for
$\Delta_1=\Delta_2=\Delta_3=\Delta$.

As a conclusion we have demonstrated how $N$-point function can be
obtained from gravity description of Galilean CFT where we have seen
that the main role plays by the base $AdS_2$ geometry.

Actually the procedure may be summarized as follow. In order to
obtain the physical correlation functions one needs to use the
standard AdS/CFT correspondence for $AdS_2$ part, though the action
of the corresponding propagating fields in the $AdS_2$ geometry gets
a contribution from the fiber  via the measure of the integral.
Otherwise the procedure follows the same as that in the standard
AdS/CFT correspondence applied for $AdS_2$.

The above procedure may be generalized to arbitrary $n$. The only
difference is that the $AdS_2$ has to be replaced by $AdS_{n+2}$. In
other words in this case the action is given by \be I=\int dt dz
d^ny_\alpha\;
z^{d-n-1}\sqrt{g}\;\frac{1}{2}(g^{ab}\partial_a\phi\partial_b\phi+m^2\phi^2)
\ee where $g_{ab}$ is the metric of $AdS_{n+2}$ given by
\eqref{AdSn}. Note that the propagating fields are subject to the
constraint $\partial_i^2\phi=0$ for $i=1,\cdots, d-n-1$.

It is important to note that the decoupling  of the fiber is due to
the particular form of the constraint. If we change the constraint
(for example by breaking the conformal symmetry via heating up the
theory) the situation may be changed.

\section{Discussions}

In this paper we have considered different contractions of a $d+1$
dimensional relativistic conformal algebra. The contraction is
defined by the scaling \eqref{scale} in the limit of $\epsilon
\rightarrow 0$. In other words if we define the velocity of $l$ th
direction as \be
v_\alpha=\frac{y_\alpha}{t},\;\;\;\;\;\;\;\;v_i=\frac{x_i}{t},\;\;\;\;\;\alpha=1,\cdots,n,\;\;i=n+1,\cdots,d,
\ee in this limit one has $v_i\rightarrow 0$. Therefore the
contraction may be thought of as taking non-relativistic limit of
the relativistic conformal algebra. In particular when $n=0$ the
resultant algebra is the Galilean conformal algebra
\cite{Gopakumar}. For $n\geq 1$ we are taking non-relativistic limit
in some directions while the others remain unchanged. So, it may be
treated as a semi-Galilean conformal algebra.

In general by a contraction  the conformal algebra in $d+1$
dimension, $so(2,d+1)$, reduces to an algebra which contains an
$so(2,n+1)\times so(d-n-1)$ subalgebra. For $n=0$ and $n=1$ the
obtained algebras have $SL(2,R)$ and $SL(2,R)\times SL(2,R)$
subalgebra, respectively. Due to this property the corresponding
algebras have infinite dimensional extension where the $SL(2,R)$'s
extend to the Virasoro algebra. Having had the Virasoro algebra in
the cases of $n=0,1$, it would be interesting to see if the (semi)
Galilean conformal algebra allows a central extension to its
Virasoro subalgebra.

Following the AdS/CFT correspondence one may suspect that the (semi)
Galilean CFTs may have dual gravity descriptions. If so, the
corresponding gravity dual should contain a factor of $AdS_{n+2}$ to
support the symmetry group $SO(2,n+1)$. Moreover to have the
symmetry group $SO(d-n-1)$ the gravity dual should also have a
factor of $d-n-1$ dimensional flat space, ${\cal M}_{d-n-1}$. On the
other hand since the semi-Galilean conformal algebra cannot be
factorized as $so(2,n+1)\times so(d-n-1)$ the bulk geometry is not a
direct product of these two spaces, though locally it may be thought
of as $AdS_{n+2}\times {\cal M}_{d-n-1}$. In fact it was argued in
\cite{Gopakumar} that at least for $n=0$ the geometry is a $d-1$
dimensional spatial space fibered over an $AdS_2$. From our
consideration we expect that in general the bulk geometry is a
$d-n-1$ dimensional spatial space fibered over an $AdS_{n+2}$. Note,
that, the corresponding gravity is given in terms of Newton-Cartan
like description when the role of time is replaced by an
$AdS_{n+2}$.

Using this picture it is easy to understand why the cases of $n=0$
and 1 have infinite dimensional extension while the other cases are
finite dimensional. In fact the reason is due to the asymptotic
symmetry of AdS space; while for $AdS_2$ and $AdS_3$ it is infinite
dimensional, for the others it is finite dimensional\footnote{It
should be compared with Schr\"odinger algebra which has infinite
dimensional extension in any dimension.}.

We have also explored the AdS/CFT correspondence for (semi) Galilean
CFTs where we have seen that the essential role is played by the
base $AdS_{n+2}$ geometry. In fact the correlation functions of the
(semi) Galilean CFTs can be evaluated by making use of propagating
fields on $AdS_{n+2}$ with proper boundary conditions and a modified
measure due to the contribution of the fiber.

An interesting application of this contraction would be to apply the
procedure to ${\cal N}=4$ four dimensional SYM theory whose gravity
dual is given by type IIB string theory on $AdS_5\times S^5$. Taking
the limit from both sides of the duality one may single out a subset
of ${\cal N}=4$ four dimensional SYM theory which has (semi)
Galilean conformal symmetry. This might give a new insight about the
AdS/CFT correspondence following \cite{BMN}.

In the context of AdS/CFT duality it is known that heating up the
dual field theory generically corresponds to adding a black hole in
the bulk gravity. Therefore we would expect that applying the above
limit one may find a gravity dual to the (semi) Galilean CFT at
finite temperature. In this case the bulk gravity background may be
interpreted as a $d-n-1$ dimensional spatial space fibered over a
base which is given by a black hole in $AdS_{n+2}$ space. It is
worth noting that in this case the contraction may by supplemented
by a shift in $\partial_t$. In this case the constraint does not
decouple from the equation of motion of propagating modes in the
base $AdS_{n+2}$ space. As a result the correlation function will
depend on the fiber coordinates too\cite{progress}.

\section*{Acknowledgments}
We would also like to thank  Hamid Afshar, Amin Akhavan, Davod
Allahbakhsi, Reza Fareghbal, Amir E. Mosaffa and Shahin Rouhani for
discussions on the different aspects of non-relativistic AdS/CFT
correspondence. This work is supported in part by Iranian TWAS
chapter at ISMO.


\end{document}